\documentclass{article}
\usepackage{dcase2024,amsmath,graphicx,url,times,booktabs, tabularx,siunitx}
\title{Machine listening in a neonatal intensive care unit}

\name{Modan Tailleur$^{1}\sthanks{Thanks to ANR AIBY4 (ANR-20-THIA-0011) for funding.}$,
      Vincent Lostanlen$^{1}\sthanks{Thanks to ANR MuReNN (ANR-23-CE23-0007-01) for funding.}$,
      Jean-Philippe Rivière$^{1}$, 
      Pierre Aumond$^{2}$
      }
\address{$^1$ Nantes Université, École Centrale Nantes, CNRS, LS2N, UMR 6004, F-44000 Nantes, France\\          
        $^2$ Université Gustave Eiffel, CEREMA, UMRAE, F-44344 Bouguenais, France
 }

\begin{document}

\ninept
\maketitle

\begin{sloppy}

\begin{abstract}
Oxygenators, alarm devices, and footsteps are some of the most common sound sources in a hospital.
Detecting them has scientific value for environmental psychology but comes with challenges of its own: namely, privacy preservation and limited labeled data.
In this paper, we address these two challenges via a combination of edge computing and cloud computing.
For privacy preservation, we have designed an acoustic sensor which computes third-octave spectrograms on the fly instead of recording audio waveforms.
For sample-efficient machine learning, we have repurposed a pretrained audio neural network (PANN) via spectral transcoding and label space adaptation.
A small-scale study in a neonatological intensive care unit (NICU) confirms that the time series of detected events align with another modality of measurement: i.e., electronic badges for parents and healthcare professionals.
Hence, this paper demonstrates the feasibility of polyphonic machine listening in a hospital ward while guaranteeing privacy by design.
\end{abstract}


\section{Introduction}
\label{sec:intro}

\begin{figure}
    \centering
    \includegraphics[width=0.8\linewidth]{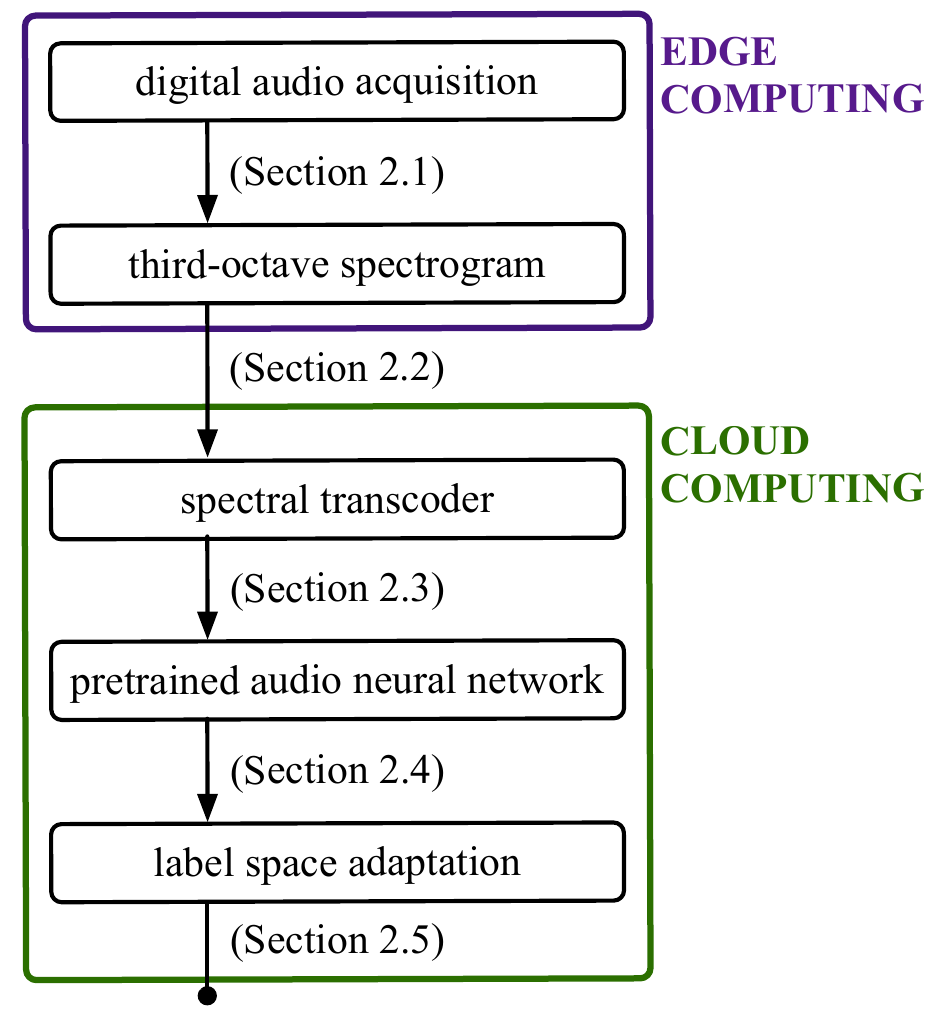}
    \caption{Flowchart of stages in the proposed approach. The first two stages are performed ``on the edge''. The last three stages are performed ``on the cloud'', i.e., on a central server.}
    \label{fig:flowchart}
\end{figure}

Sound is a reliable and non-invasive carrier of information about human health \cite{beach2014medical}.
Yet, historically, the subfield of medical acoustics has mainly focused on analyzing sounds as \emph{produced} by patients: stutter \cite{sheikh2022machine}, crackles \cite{pramono2017automatic}, cough \cite{alqudaihi2024cough}, and so on.
Much less is known about the sounds as \emph{heard} by patients in a clinical setting: as experimental psychologists have pointed out, the detailed description of acoustic events in intensive care units (ICU's) is typically overlooked in favor of sound pressure level measurements (SPL) \cite{mackrill2013experiencing}.
Meanwhile, exposure to anthropogenic noise at unsafe SPL levels is known to induce stress, cognitive impairment and sleep disorders in children \cite{bronzaft2021supporting} and adults \cite{murphy2022environmental}, thus calling for urgent remediation.

The case of neonatal intensive care units (NICU's), where premature babies receive special care to grow and survive, presents an even greater gap in research than adult ICU's \cite{lenzi2023improving}.
During their time in the NICU, preterm infants are exposed to unpredictable sensory stimuli while undergoing a protracted period of rapid brain growth, causing lasting effects on cognitive ability \cite{gray2004effects}.
Unfortunately, the auditory physiology and  cognition of neonates have received insufficient attention from scientists until recently \cite{philbin2017sound}.

What is known with certainty is that parents have an essential role to play in the development of their newborn babies  \cite{baley2015skin}.
Indeed, an approach sometimes described as ``kangaroo care'' involves prolonged periods of skin-to-skin contact between the baby and either of the two parents, in addition to incubator placement.
Promoting this approach requires to take the well-being of parents into consideration so that they feel included into collective care work.

For this purpose, we have launched a project on ``listening to family experiences in the neonatological ward'', or LIFEWARD for short.
Here, the word ``listening'' is understood as both qualitative and quantitative: i.e., as enacted by interviews with parents as well as autonomous acoustic sensors.
Although there is scientific consensus around the value of semi-structured interviews for neonatology---see, for example, \cite{medina2018bonding}---the same cannot be said about machine listening.
This is for at least three reasons.
First, the deployment of acoustic sensors in a hospital raises pressing concerns about privacy preservation and cybersecurity.
Secondly, the application of machine learning to the NICU is not straightforward, for lack of annotated training data.
Thirdly, and perhaps most fundamentally, machine listening systems have not yet demonstrated their ability to reconstruct objective information about social bonds in the NICU.
Addressing these three challenges is necessary before envisioning the integration of machine listening instruments within the toolkit of patient experience research.

In this article, we present a proof of feasibility of machine listening for neonatology.
Prior work in this domain has focused on a single class of sound event---namely, the spontaneous cries of preterm newborns \cite{cabon2021automatic}.
Meanwhile, our system is a multilabel sound event detector for adult voices, footsteps, oxygenators, and alarm devices.
Furthermore, the originality of our approach is that it integrates all aspects of machine listening, from digital audio acquisition to sound event detection, into a mixed pipeline that involves both edge computing and cloud computing.

Figure \ref{fig:flowchart} decomposes our approach into five stages, described in Sections 2.1 through 2.5.
To comply with standards of privacy and security, the LIFEWARD sensor does not store the acquired waveforms.
Rather, it extracts a third-octave spectrogram on the fly; i.e., a coarse estimate of power spectral density over  windows of duration \SI{125}{\milli\second}.
This, in turn, brings its own challenges for sound event detection, which typically requires finer spectrotemporal information.
We address this challenge via a pretrained ``spectral transcoder''; i.e., a deep neural network for nonuniform resampling the time--frequency domain.
We pass the output of the spectral transcoder to a pretrained audio neural network (PANN), with AudioSet as its label space.
Lastly, we use domain-specific knowledge to narrow down this taxonomy for the NICU.

Section 3 presents the result of an in-progress study at an NICU, and provides tentative evidence for the feasibility of the proposed approach.
Indeed, neural network predictions appear to coincide with isolated sound events of interest (Section 3.1) as well as timestamps from a non-audio modality of human presence (Section 3.2).

\section{Methods}

\subsection{Acoustic sensor}
Our acoustic sensor is a Raspberry Pi, inspired by previous work on urban noise monitoring \cite{ardouin2018innovative}.
It acquires audio from an external USB microphone, specifically, the micW i436. 
The i436 is an omnidirectional electret microphone with a capsule diameter of approximately 7 mm, in compliance with NF EN 61672 Class-2 standards.
Its sensitivity and frequency response has been calibrated manually by the manufacturer.
After digital--analog conversion, the sample rate is 32 kHz.
Our sensor is powered by the grid and ``air-gapped'', i.e., physically isolated from the public Internet and from each other.
This is to reduce the risk of malicious data access.

\subsection{Third-octave spectrogram}
We use fast Fourier transforms (FFT) to design a third-octave filterbank with bands ranging from 20 Hz to 12.5 kHz, in compliance with the ANSI S1.1-1986 and IEC 61260-1:2014 standards \cite{gontier2017efficient}.
We extract the magnitude response of each filter over non-overlapping subbands of duration 125 ms.
These operations are implemented in the C language, compiled for the Raspberry Pi, and executed in real time.
The result is stored incrementally on a non-volatile memory (``SD card'').

A perceptual evaluation on twelve subjects has shown that the third-octave spectrogram does not contain sufficient information to recover intelligible speech, at least via classical signal processing techniques---namely, Moore-Penrose pseudoinverse and Griffin-Lim algorithm for phase retrieval \cite{gontier2017efficient}.
Thus, the third-octave spectrogram representation can be said to be privacy-aware, in the sense it mitigates the severity of a security breach should the SD card were to be lost or stolen in the healthcare facility.

Another advantage of computing third-octave spectrograms on the edge resides in its bitrate: around 3.71 kilobytes per second (kbps).
This is lower than MP3 (128--320 kbps) and lossless audio (around 1 Mbps).
The bitrate of third-octave spectrograms translates to around 320 megabytes per day, or 117 gigabytes per year.
Thus, a single SD card suffices to contain all the spectrogram data over a longitudinal survey spanning the full length of stay of the preterm infant at the NICU.

\subsection{Spectral transcoder}
Previous work in urban environments has shown the potential of the third-octave spectrogram as a feature for sound event classification, both in supervised and self-supervised scenarios \cite{gontier2021polyphonic}.
Yet, this previous work is unapplicable in the context of the NICU, for lack of annotated training data.
Furthermore, note that it would not be possible to launch our own annotation campaign because, as explained before, our sensors do not record audio.
We propose to circumvent this problem by relying on a pretrained audio neural network (PANN) for multilabel sound event detection and classification \cite{kong2020panns}.

Here, a second issue arises: PANN does not operate upon the third-octave spectrogram but on a mel-frequency spectrogram, which has a finer temporal resolution (hop size of 10 ms) and a finer spectral resolution (64 bins on the mel scale).
In principle, the required change of resolution could be achieved by a linear non-uniform resampler.
Yet, in practice, this produces a blurry time--frequency representation which is not recognized by PANN as containing any events of interest.
Against this issue, a deep neural network was developed by Tailleur et al. \cite{tailleur2023spectral}, which we call \emph{spectral transcoder}, so as to recover a plausible mel-frequency spectrogram from a third-octave spectrogram measurement.

The spectral transcoder is a convnet with six layers. It is trained on TAU Urban Acoustic Scenes 2020 Mobile dataset \cite{mesaros2018multi} in a ``teacher--student'' scenario. The teacher is the composition of mel-frequency spectrogram and PANN whereas the student is the composition of third-octave spectrogram, spectral transcoder, and PANN.
In other words, the spectral transcoder is not trained to minimize its mean square error with the mel-frequency ground truth (as a linear model would) but to generate a mel-frequency spectrogram whose spectrotemporal content has the same distribution of sound events as the ground truth. The training process involves minimizing a binary cross-entropy loss, computed between the PANN output of the student and that of the teacher, by updating solely the transcoder's parameters. This is a kind of super-resolution procedure in which the implicit knowledge about the spectrotemporal characteristics of natural audio sounds is distilled from PANN into the spectral transcoder under the form of convnet weights.
We refer to  \cite{tailleur2023spectral} for more details on the spectral transcoder.

\subsection{AudioSet classification with PANN}
Our PANN of choice is a residual network with 38 layers, or ResNet38 for short.
It contains around 74M parameters.
To this day, it is regarded as one of the most accurate general-purpose multilabel audio classifier among those which take the mel-frequency spectrogram as input.
The PANN is trained on AudioSet, a dataset which contains over 2M 10-second audio clips which were extracted from YouTube videos.
In the next sections, we refer to the composition of pretrained spectral transcoder and PANN as ``PANN-1/3oct'' model.
We refer to \cite{kong2020panns} for further details on PANN.

Since PANN is a multilabel classifier, its output vector is unnormalized.
For the sake of visualization, we have found it beneficial to rank predictions in decreasing order, normalize rank by the number of classes, and apply an inverse power transform.
Given predictions $\boldsymbol{x}[k]$ for each class $k$, this procedure yields the $\alpha$-compressed reciprocal rank
\begin{equation}
\boldsymbol{y}[k] = \left(\dfrac{1}{\sigma^{-1}[k]}\right)^{\alpha},
\end{equation}
where $k$ is the class index, $\sigma$ is the sorting permutation such that $(\boldsymbol{x}\circ\sigma)[1] > \ldots > (\boldsymbol{x}\circ\sigma)[K]$, and $\alpha<1$ is a constant exponent.
We set $\alpha=0.5$ in this paper.

\subsection{Label space adaptation}
The PANN-1/3oct model analyzes a third-octave spectrogram snippet of duration equal to 10 seconds and returns a vector of dimension 527, corresponding to the classes in AudioSet dataset.
These classes are a subset of the 623-class AudioSet ontology\footnote{Link to complete list of classes in the AudioSet dataset: \\\url{https://research.google.com/audioset/dataset/index.html}}, which has been defined by a Google Research team after scraping web-scale text data for ``Hearst patterns'', i.e., of either of these forms \cite{gemmeke2017audio}:
\begin{quote}
    [...] sounds such as X or Y [...]
    
    [...] X, Y, and other sounds [...]
\end{quote}
This approach has proven fruitful for general-purpose audio classification: we refer to \cite{pellegrini2023adapting} for a review.
Yet, it is unsuitable for the ICU, whose distribution of sound events is inadequately represented by textual mentions of sound events on the web.
At the same time, training a classifier from scratch on a new taxonomy is out of the question for reasons of privacy preservation, as explained earlier.

Instead, we simply run the PANN-1/3oct model on third-octave spectrogram data from the NICU and look for some frequently occuring AudioSet classes.
We find four activity patterns of interest: ``conversation'', ``walk, footsteps'', ``train'', and ``electronic music''.
Although the former two sound events are plausible, the latter two are clearly not.
Yet, after interviewing NICU employees, we may hypothesize that they yield indirect information: i.e., that ``train'' actually corresponds to the rumble of the oxygenator while ``electronic music'' corresponds to the ringtone of the hospital phone.
We summarize this correspondence in Table \ref{tab:label_mapping}.

\begin{table}[h]
    \centering
    \begin{tabular}{l l}
        \toprule
        \textbf{Neonatal Intensive Care Unit (NICU)} & \textbf{AudioSet} \\
        \midrule
        Conversation & Conversation \\
        Footsteps & Walk, footsteps \\
        Oxygenator & Train \\
        Hospital phone & Electronic music \\
        \bottomrule
    \end{tabular}
    \caption{Mapping of sound event labels from the neonatal intensive care unit (NICU) to AudioSet.}
    \label{tab:label_mapping}
\end{table}

\section{Application}
\subsection{Deployment in a neonatal intensive care unit}
Since 2018, a design company\footnote{Sensipode} have been partnering with Nantes University hospital and a nonprofit organization\footnote{the B.E.R.S.E association} to enhance the inclusion of parents in the NICU.
The nonprofit organization collaborated with designers to refurnish a care room so as to facilitate the presence of parents alongside their newborn.
In this context, the LIFEWARD sensor has offered the necessary guarantees for a safe and privacy-aware deployment in the NICU.
We have obtained the approval of an ethical review board to deploy this sensor\footnote{Groupe Nantais d'Éthique dans le Domaine de la Santé (GNEDS) - n°22-09-090}.
Six families have given their informed consent to participate in the LIFEWARD study: three in the aforementioned redesigned room and three in a standard room. 
The length of stay is approximately 90 days for each family.
Thus, we have collected third-octave spectrogram data over 18 cumulated months.

\subsection{Visualization of sound events }
We now collect a few waveform-domain samples from the sound events of interest in a real NICU environment.
This data collection stage is carried out with a handheld device, over short durations, and with the collaboration of NICU professionals.
Specifically, we ring various kinds of alarms, activate oxygenators and other pumps, stomp our feet, and so forth.
Admittedly, these sounds are too few to offer an independent quantitative evaluation of PANN-1/3oct: we refer to \cite{tailleur2023spectral} for that matter.
Still, they may serve as suggestive evidence for the fact that the correspondences which we hypothesized in Table 1 are adequate and useful in practice.

Figure \ref{fig:spectrograms} illustrates our findings for each of the four classes of interest.
For example, we notice vertical patterns of high energy in the recording of footsteps, versus horizontal patterns in the recordings of oxygenator.
These simple observations corroborate the prediction of the PANN-1/3oct model with label space adaptation: see Table  \ref{tab:label_mapping}. Those examples demonstrate one of the advantages of using the transcoder: one can double-check model predictions by displaying the transcoded spectrogram, despite not being able to listen to the underlying audio waveform.

\begin{figure*}[h!]
    \centering\includegraphics[width=\linewidth]{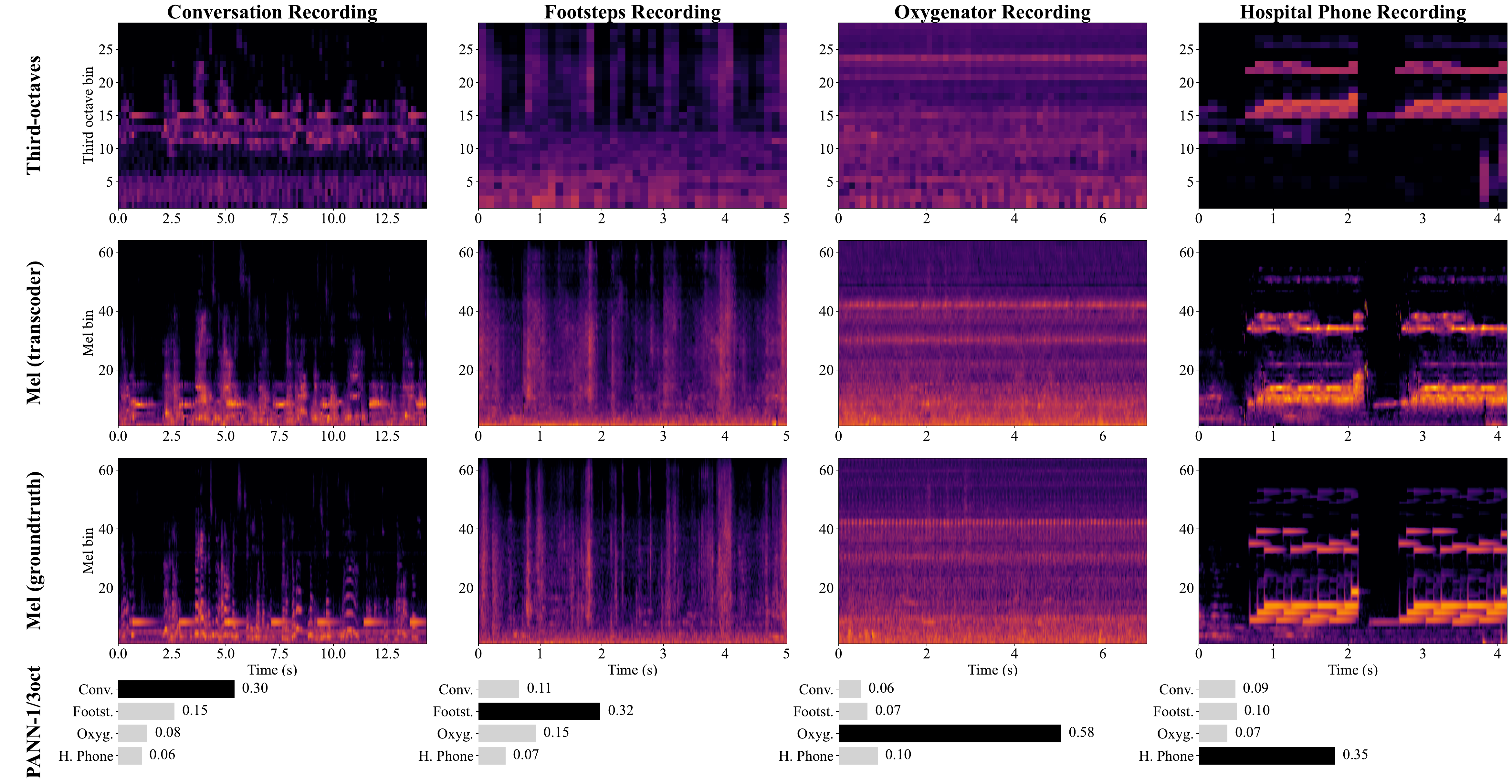}
    \caption{Spectrograms from audio recordings in the Neonatal Intensive Care Unit (NICU). First row corresponds to audio recordings transformed into fast third-octaves spectrograms. Second row corresponds to Mel spectrograms transcoded with the transcoder. Third row corresponds to groundtruth Mel spectrograms, obtained with Mel transformation on the waveform. PANN-1/3oct predictions, using the mapping between Audioset classes and NICU classes, are shown in fourth row.}
    \label{fig:spectrograms}
\end{figure*}

\subsection{Proof of feasibility for continuous monitoring}
The previous section has confirmed the interest of the PANN-1/3oct model in the context of isolated sounds from the NICU, as acquired by a handheld device.
It remains to be seen if this model remains informative in a real-world polyphonic context, as acquired by the LIFEWARD sensor.
For this purpose, we propose to compare the detected events with another modality of measurement: i.e., electronic badges worn by parents and healthcare professionals.
Via near-field communication (NFC), these badges yield information about who is present in the care room at any given time.
Hence, they offer indirect confirmation for the feasibility of machine listening in the NICU, while remaining non-invasive and privacy-aware.

Figure \ref{fig:enter-label} shows an example of PANN-1/3oct predictions from our real-world NICU dataset, together with timestamps from electronic badges.
We notice that segments during which two adults are present in the room coincide with a rise in the presence of conversation---and, to a lesser degree, of footsteps.
Meanwhile, the lowest values for the ``conversation'' class correspond to segments in which only one adult is present in the room.
Yet, we recognize that these are only anecdotal observations.
Future research is needed to expand the comparison of acoustical and non-acoustical information to a larger scale; i.e., multiple days and multiple rooms.


\begin{figure*}[h!]
    \centering
    \includegraphics[width=\linewidth]{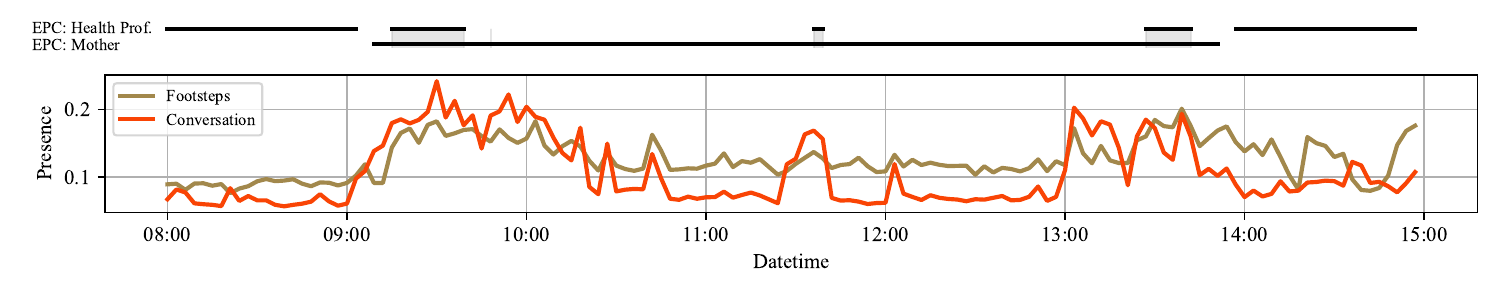}
    \caption{Presence of conversation and footsteps on a day of April 2023 in one room, as averaged over three-minute intervals.
    The badge of the health professional (EPC) and of the mother are also shown during the period. The shaded areas denote intervals in which more than one adult is present in the room.}
    \label{fig:enter-label}
\end{figure*}

\section{Conclusion}
The DCASE community has a key role to play at the intersection between sound design and healthcare.
Yet, fulfilling this role comes with challenge of its own, such as: privacy, cybersecurity, and limited labeled data.
In this article, we have presented a first prototype of acoustic sensor which demonstrates the feasibility of sound event detection in a neonatal intensive care unit (NICU).
The main limitation of our study is that, because our sensor does not record audio as waveforms, it is impossible to establish a ``ground truth'' by expert annotation.
We have circumvented this limitation in two way: first, by evaluating the system on well-controlled isolated sounds; and second, by matching the sequence of detected sound events with non-acoustical information.
In the future, we plan to refine the integration of multiple data modalities towards a more comprehensive understanding of patients and their lived experiences.

\section{ACKNOWLEDGMENT}
\label{sec:ack}

M.T.\  and V.L.\ thank Khrystyna Povkh for administrative management.
V.L.\ thanks Sandie Cabon, Mathieu Lagrange, and Florence Levé for their insights.
JP.R.\ thanks Ouest Industries Créatives, for their financial support (grant: OIC-2021-PEEL); and to Sensipode, the B.E.R.S.E. association, the STid Group, Didier Poiraud, Nicolas Fortin, Dr. Jean-Baptiste Muller, Judikaëlle Jacquin, Yannick Prié, as well as all the patients and hospital staff members for their valuable insights.


\bibliographystyle{IEEEtran}
\bibliography{refs}

%
%
%
%
%
%
%
%
%

\end{sloppy}
\end{document}